\documentclass[aps,amssymb,amsmath,amsfonts,prl,superscriptaddress,twocolumn]{revtex4}
\usepackage[utf8]{inputenc}
\usepackage{amsmath} 
\usepackage{graphicx}
\usepackage[usenames]{color} 
\usepackage{amssymb} 
\usepackage[toc,page,titletoc]{appendix}

\newcommand{\beq}{\begin{equation}}
\newcommand{\eeq}{\end{equation}}
\def\ba{\begin{eqnarray}}
\def\ea{\end{eqnarray}}

\newcommand{\ket}{|\psi\ra}

\def\ra{{\rangle}}

\newcommand{\ie}{\textit{i.e.} }
\newcommand{\eg}{\textit{e.g.} }

\begin{document}
	
\setlength\parskip{1.0mm}

\title{Quantum Singular Value Decomposer}

\author{Carlos Bravo-Prieto}
\affiliation{Barcelona Supercomputing Center, Barcelona, Spain.}
\affiliation{Departament de F\'isica Qu\`antica i Astrof\'isica and Institut de Ci\`encies del Cosmos (ICCUB), Universitat de Barcelona, Barcelona, Spain.}
\author{Diego García-Martín}
\affiliation{Barcelona Supercomputing Center, Barcelona, Spain.}
\affiliation{Departament de F\'isica Qu\`antica i Astrof\'isica and Institut de Ci\`encies del Cosmos (ICCUB), Universitat de Barcelona, Barcelona, Spain.}
\affiliation{Instituto de F\'{i}sica Teórica, UAM-CSIC, Madrid, Spain.}
\author{Jos\'{e} I. Latorre}
\affiliation{Departament de F\'isica Qu\`antica i Astrof\'isica and Institut de Ci\`encies del Cosmos (ICCUB), Universitat de Barcelona, Barcelona, Spain.}
\affiliation{Nikhef Theory Group, Science Park 105, 1098 XG Amsterdam, The Netherlands.}
\affiliation{Center for Quantum Technologies, National University of Singapore, Singapore.}

\begin{abstract}
We present a variational quantum circuit that produces the Singular Value Decomposition of a bipartite pure state. The proposed circuit, that we name Quantum Singular Value Decomposer or QSVD, is made of two unitaries respectively acting on each part of the system. The key idea of the algorithm is to train this circuit so that the final state displays {\sl exact output coincidence} from both subsystems for every measurement in the computational basis. Such circuit preserves entanglement between the parties and acts as a diagonalizer that delivers the eigenvalues of the Schmidt decomposition. Our algorithm only requires measurements in one single setting, in striking contrast to the $3^n$ settings required by state tomography. Furthermore, the adjoints of the unitaries making the circuit are used to create the eigenvectors of the decomposition up to a global phase. Some further applications of QSVD are readily obtained. The proposed QSVD circuit allows to construct a SWAP between the two parties of the system without the need of any quantum gate communicating them. We also show that a circuit made with QSVD and CNOTs acts as an encoder of information of the original state onto one of its parties. This idea can be reversed and used to create random states with a precise entanglement structure.

\end{abstract}

\maketitle
\setlength\parskip{1.5mm}

\bigskip
\noindent{\sl \bf Introduction}
Much progress has been made towards a better understanding of bipartite and multipartite entanglement of quantum systems in the last decades. Among the many figures of merit that have been put forward to quantify entanglement, the von Neumann entropy stands out as it finely reveals the quantum correlations between subparts of the system. Yet, the explicit computation of this entropy, as well as many other bipartite measures of entanglement, relies on a clever decomposition of the tensor that describes a two-party system. On the experimental side, although entropies remain elusive as no direct observable describes them in a straight way, a few approaches have been proposed in ultracold atoms \cite{exp_entropy1, exp_entropy2}.

The fundamental mathematical tool to analyze bipartite entanglement is the so called Schmidt decomposition \cite{schmidt,peres}, also named as Singular Value Decomposition (SVD). Given the knowledge of a bipartite state in its tensor form, the SVD casts this tensor onto a simpler diagonal form, which unveils the entanglement structure of the original state. In practise, the SVD requires the knowledge of the coefficients of the state and needs further computational effort to get the eigenvalues and eigenvectors that fully characterize the state.

Two proposals have been put forward for diagonalizing a matrix on a quantum computer. One of them relies on exponentiation of the matrix and subsequent application of the Quantum Phase Estimation procedure \cite{Lloyd}. The second proposal \cite{VQSD} is a variational algorithm that seeks to directly diagonalize a density matrix $\rho$ by simultaneously acting on two copies of the quantum state described by $\rho$. The cost function to be minimized in this case quantifies how far the state is from being diagonal in terms of purity. There also exist quantum algorithms \cite{entspec, alamosent} that compute Rényi entropies, and from them, the largest eigenvalues of reduced density matrices. Finally, a different approach using a continuous-variable quantum computer is considered in \cite{cont}.

We shall here present a quantum circuit that produces the elements of the SVD of a pure bipartite state, that we shall call QSVD for Quantum Singular Value Decomposer. As we shall see, the circuit we propose is made of two unitaries, each acting on a separate subpart of the system, that can be determined in a variational way. The frequencies of the outputs of the final state in the circuit deliver the eigenvalues of the decomposition without further treatment. Also, the eigenvectors of the decomposition can be recreated from direct action of the adjoint of the unitaries that conform the system on trivial states.

The key ingredient of the algorithm is to train the circuit on exact coincidence of outputs. This is a subtle way to force a diagonal form onto the state. It also provides an example of a quantum circuit which is not trained to minimize some energy, but rather to achieve a precise relation between the superposition terms in the state (other examples can be found in \cite{VQSD, autoencoder, carolan2020, VQLS, tangle}). We further verify the QSVD algorithm on simulations. 

A peculiar bonus of our approach is that the QSVD provides a means to perform a SWAP between parties without ever having quantum communication between them. Another one is that the QSVD can be turned into an encoder of quantum information.

Our proposal is a hybrid classical-quantum algorithm, much in the spirit of recent developments in the field of Quantum Computation for the Noisy Intermediate-Scale Quantum  (NISQ) era \cite{nisq,variational}. This means that the basic circuits may be shallow (accuracy can be increased by increasing the depth), and therefore amenable to implementation on near term quantum computers without error correction.

\bigskip
\noindent{\sl \bf Classical Singular Value Decomposition}
The Singular Value Decomposition is a powerful mathematical technique which is ubiquitously used to analyze tensors with two indices. It simply says that any such tensor can be cast onto a diagonal form using two unitary matrices that act on each of its indices.

Let us briefly review how the SVD is computed. Consider a bipartite pure state $\ket_{AB} \in \mathcal{H}_A\otimes \mathcal{H}_B$, 
\beq  \label{vector} \ket_{AB} = \sum_{i=1}^{d_A} \sum_{j=1}^{d_B}
 c_{ij}\,|e_i\ra_A|e_j\ra_B\,, \eeq
where $d_{A,B}$ are the dimensions of the susbsystems Hilbert spaces $\mathcal{H}_{A,B\,}$, $\{|e_k\ra_{A,B}\}$ are the computational-basis states in $\mathcal{H}_{A,B\,}$, and the complex coefficients $c_{ij}$ obey a normalization relation. This state can be written in its Schmidt form,
 \beq \ket_{AB} = \sum_{i=1}^\chi \lambda_i\,|u_i\ra_A |v_i\ra_B\,, \eeq
where $\chi$ is the Schmidt rank (\ie the number of Schmidt coefficients different from zero), which is always equal or smaller  than the minimum of $d_A$ and $d_B$; $\lambda_i$ are real positive eigenvalues that can be sorted in decreasing order, and $\{|u_i\ra_A\}$ and $\{|v_i\ra_B\}$ form a orthonormal basis for subsystems $A$ and $B$ respectively. 

The analytical way to find the SVD of a given vector (\ref{vector}) needs to start from the tensor $c_{ij}$, then compute the reduced density matrix for each subsystem,
$\rho_A=Tr_B |\psi \rangle_{AB} \langle \psi |$ and $\rho_B=Tr_A |\psi \rangle_{AB} \langle \psi |$,
and then perform two diagonalizations,
$\rho_A=\sum_{i=1}^\chi\lambda_i^2 |u_i\rangle_A \langle u_i|$ and 
$\rho_B=\sum_{i=1}^\chi\lambda_i^2 |v_i\rangle_B \langle v_i|$.
As a result, the original vector can be cast in the basis of the eigenvectors 
$ \{|u_i\rangle_A\}$ and $\{|v_i\rangle_B \}$ of both diagonalizations that share the same eigenvalues. 
The sign of each $\lambda_i$ can be taken positive as a phase can always be absorbed into either $|u_i\rangle_A$ or $|v_i\rangle_B$.

Note that the Schmidt rank $\chi$ is in itself a first measure of entanglement. Furthermore, the usefulness of the SVD can be illustrated by computing the von Neumann entropy $S$ of this state for the $A$-$B$ bipartition: $S=-\rm{Tr }\left(\rho_A \log\rho_A\right) =  -\rm{Tr }\left(\rho_B \log\rho_B\right)=-\sum_{i=1}^\chi \lambda_i^2 \log \lambda_i^2$. It also follows that all R\'enyi entropies can be computed once the eigenvalues of the SVD are known \cite{renyi}.

The classical construction of the SVD can only be used on an actual quantum state after performing its full tomography. Indeed, for a state made out of $n$ qubits it is necessary to perform measurements in $3^n$ different settings (\ie one for each non-commuting combination of tensor products of Pauli operators $\{\sigma_x, \sigma_y, \sigma_z\}$) to reconstruct the original tensor $c_{ij}$ \cite{tomography}, and then perform all the computations sketched above. Furthermore, the classical computation of the entropy to be performed from the tensor describing the state may get out of reach for large systems, since it scales  exponentially with the number of qubits.

\bigskip
\noindent{\sl \bf Circuit for Quantum Singular Value Decomposer (QSVD)}
Here we present a novel way to compute the eigenvalues and obtain the physical eigenvectors of the SVD of a pure state $\ket_{AB}$ using a quantum circuit, that we shall name QSVD for Quantum Singular Value Decomposer. Our technique needs copies of the original state.

The key idea of our method is to find a circuit that provides the following transformation of the original state:
\beq 
U_A\otimes V_B\,\ket_{AB} = \sum_{i=1}^\chi \lambda_i\,e^{i\alpha_i}|e_i\ra_A|e_i\ra_B\,,
\eeq
where $U_A|u_i\ra_A=e^{i\beta_i}|e_i\ra_A$ and $V_B|v_i\ra_B=e^{i\gamma_i}|e_i\ra_B$, with $\alpha_i=\beta_i+\gamma_i \in[0,2\pi)$; the $i$ in the exponent is the imaginary unit.

\begin{figure}
	\centering
	\includegraphics[width=7.9cm,height=5.5cm]{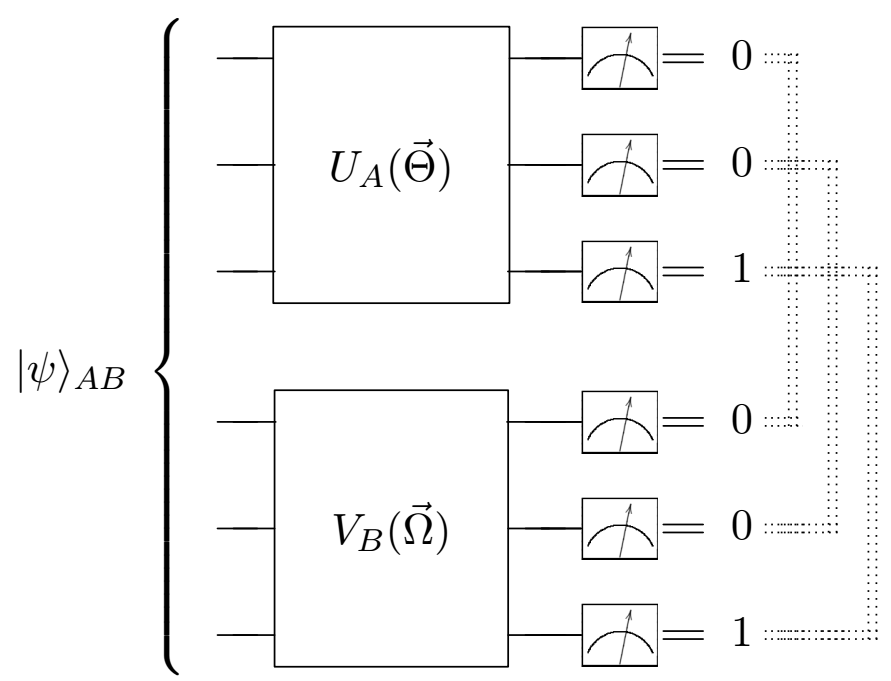}	
	\caption{Parametrized unitary transformations implementing the Quantum Singular Value Decomposer (QSVD). Training is based on demanding exact output coincidence for both parties and for every measurement.}
	\label{fig:QSVD}
\end{figure}

The way to find the desired circuit emerges from the following observation. Given that the new Schmidt bases for the two subsystems are right away the computational-basis vectors (up to individual global phases), each time we perform a measurement we should find {\sl exact output coincidence} between the respective observations in $A$ and $B$ (FIG. \ref{fig:QSVD}). Let us consider the example of two subspaces of two qubits. Then, if the result in the first subsystem turns out to be \eg 00, the result in the second subsystem should also be 00. There is always a pair of unitaries $U_A$ and $V_B$ achieving this exact output coincidence, since they simply correspond to a change of basis from the Schmidt eigenvectors. 

Note that the entanglement spectrum between $A$ and $B$ (\ie the eigenvalues $\{\lambda_i\}$ of the reduced density matrices $\rho_A,\rho_B$) has not changed, nor has the von Neumann entropy. This allows to obtain an estimation of the Schmidt coefficients $\{\lambda_i\}$, which will just be the observed, normalized probabilities for each possible {\sl coincident} outcome for the two subsystems, obtained from repeated preparation of the state, application of the QSVD and measurement. In turn, these coefficients provide several entanglement figures of merit, such as the von Neumann entropy.

Once the two unitaries $U_A$ and $V_B$ have been obtained, it is now possible to reconstruct the vectors that would be needed in the original SVD, up to a complex phase. They simply correspond to
\beq \label{eigenvectors}
e^{-i\beta_i}|u_i \rangle_A = U_A^\dagger\, |e_i\rangle_A\;, \qquad 
e^{-i\gamma_i}|v_i \rangle_B = V_B^\dagger \,|e_i\rangle_B \,,
\eeq
where $|e_i\rangle$ can be created by just applying X-gates onto the initial $|00...00\rangle$ state at the begining of the computation. The algorithm  has taken the original state to a very specific form, the one of exact output coincidence, to determine the unitaries, and then the adjoint of the same unitaries are used to reconstruct the eigenvectors. 
The global phases $\{\beta_i,\gamma_i\}$ are irrelevant in the characterization of the individual eigenvectors, but if one is interested in the relative phases between these vectors in the original $\ket_{AB}$, then one may need to resort to tomography.

The algorithm we have presented has some extra degrees of freedom. For the sake of clarity we explicitly demanded exact output coincidence. But, this is not necessary in a strict sense. It suffices that each unique output from subsystem $A$ is matched by some other unique output from subsystem $B$. This means that there is freedom of permutation for, say, output from $B$. Such a permutation is just another unitary on the $B$ side. However, freedom of permutation (\ie alternative training) must be avoided in two further applications of the QSVD (SWAP without quantum communication and quantum encoder), which we shall present below. Freedom of phase, in contrast, does not have any effect on them.

Another obvious comment of the algorithm is related to the possibility of having partitions with different dimensions. In such a case, the larger subsystem will have a number of irrelevant elements in its basis that will never tick on measurement.

\bigskip
\noindent{\sl \bf Variational QSVD}
The key role of the exact output coincidence is the guide to construct a quantum circuit to perform this task. Indeed, it is possible to train a variational version of the QSVD that will approximate the exact QSVD. 

We first need to construct the two needed unitaries as a quantum circuit made of entangling gates and single qubit rotations. This circuit is thus characterized by a set of classical parameters. We may choose for instance the architecture shown in FIG. \ref{fig:unitary} of Supplementary Material, where all the free parameters correspond to angles of rotation for single qubits
 $\vec{\Theta}$ and $\vec{\Omega}$ for subsystem $A$ and $B$ respectively. The variational form of the QSVD reads now
\begin{multline}
\label{VEE}
  \ket_{AB}  \;\xrightarrow{QSVD}\; U_A(\vec{\Theta})\otimes V_B(\vec{\Omega})\,\ket_{AB} \\ = \sum_{i=1}^\chi \lambda_i\, e^{i\alpha_i}|e_i\ra_A|e_i\ra_B \,.
\end{multline}

At the outset, random values for the parameters might be used and the circuit does not issue states that show exact output coincidence for all measurements. A figure of merit for the wrong answer is simply the total amount of non-coincidental measurement outcomes, which shall be minimized. In order to help accelerate convergence, different outcomes for each subsystem may be penalized by their Hamming distance, which is just the number of symbols that are different in the binary representation of the two results. Thus, the cost function $C$ to be minimized simply is
\beq \label{eq:cost_function} C\equiv \sum_j d_H(M_j^A,M_j^B) \equiv \sum_q \frac{1 - \langle\sigma_z^{q,A}\sigma_z^{q,B}\ra}{2}\,,\eeq
where $d_H$ denotes the Hamming distance and $M_j^{A,B}$ are the results of the $j$-th measurement in the computational basis for $A$ and $B$, respectively. Equivalently, it can also be defined in terms of 2-local $\sigma_z$ Pauli operators, where the index $q$ runs over all the qubits in the smallest subsystem. We may now apply machine learning techniques to find the optimal parameters that provide exact output coincidence. Notice that this cost function has a value of zero if and only if the Singular Value Decomposition is successfully completed. Note as well that it is defined in terms of 2-local observables and therefore, it does not suffer, for circuits of depth $\mathcal{O}(\log\,n)$, from the problem of exponentially vanishing gradients \cite{plateausLANL}. We emphasize that there is no need to perform any tomography, nor to involve any measurement of non-trivial observables. This simplification is related to the fact that there is no need to measure any relative phase. Therefore, the QSVD implies an exponential reduction in the number of measurement settings compared to state tomography, which requires $3^n$.

\begin{figure*}[t]
	\centering
	\includegraphics[scale=0.33]{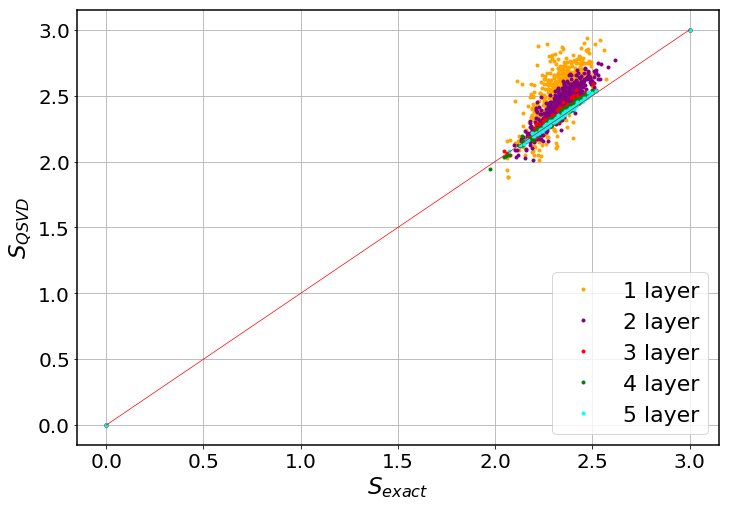}
	\hspace{0.64cm}
	\includegraphics[scale=0.33]{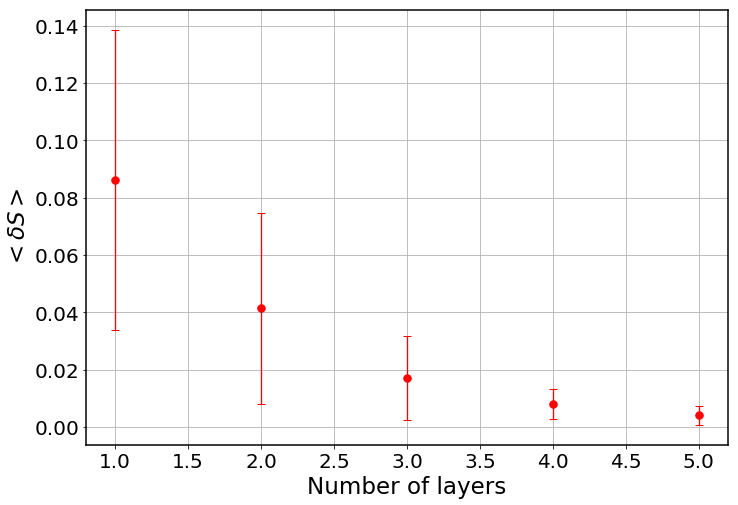}
	\caption{{\sl Left:} Von Neumann entropy computed from the variational form of QSVD {\sl vs.} exact entropy, for random states (including a product state and AME state) of 6 qubits and natural bipartition. As the number of layers increases, we observe convergence towards the exact entropy. {\sl Right:} Mean relative error in the estimation of the entropy {\sl vs.} number of layers (error bars represent the standard deviation). The error decreases exponentially with the depth of the circuit, as suggested by the Solovay-Kitaev theorem.\label{fig:entropies}}
\end{figure*}

The convergence of the method depends on two distinct elements. First, the potential convergence of a variational QSVD to the exact QSVD is controlled by the Solovay-Kitaev theorem \cite{kitaev}. Indeed, we are just looking for an approximation to a unitary using a complete set of gates. This means that there exists a quantum circuit that approximates the desired unitary with error $\delta$, \ie $|U_{exact}-U(\vec{\Theta})|<\delta$, with a number of gates $k$ that scales as $k\sim \log^c \frac{1}{\delta}$, with $1\leq c<4$, for a fixed number of qubits \cite{harrow}. In other words, the error in the unitary may potentially decrease exponentially with the depth of the circuit, for a fixed number of
qubits. This, in practice, will depend on the circuit ansatz and the success of the
optimization procedure. The number of layers of the variational circuit (see FIG. \ref{fig:unitary} of Supplementary Material) has to increase polynomially with the system syze. Under these conditions, a classical search algorithm needs only to explore a polynomial number of dimensions. Finding the optimal parameters may nonetheless encounter exponentially vanishing gradients \cite{plateaus} or local minima for deep quantum circuits, that need to be circumvented using appropriate optimization strategies \cite{platStrat, wecker, AAVQE, nikolaj, kubler2019}.

Second, the QSVD samples from a multinomial distribution. As such, the additive error for each output probability $p_i$ scales
as  $\sqrt{p_i(1-p_i)/s}$, where $s$ is the number of samples. The total number of measurements is related to the error which is aimed at, which in turn will depend on the Schmidt rank. We may then consider two different cases: (i) the Schmidt rank increases polynomially with the number of qubits, and (ii) the Schmidt rank increases exponentially with the number of qubits. In the first case (i), only a $poly(n)$ number of measurements is needed to achieve a low relative error, whereas in a worst-case scenario (ii), this number is exponential if one is to estimate all eigenvalues with a low relative error. The latter case follows naturally from the fact that we are asking for an exponential amount of information. However, many physically relevant states, \eg in condensed matter systems, do not exhibit an exponentially large Schmidt rank \cite{many-body}.

The variational approach to the QSVD can be verified on a simulation. We have considered random states with $c_{ij}=a_{ij} + i\, b_{ij}$ such that $a_{ij}$ and $b_{ij}$ are random real numbers between -0.5 and 0.5, further restricted by a global normalization. These states tend to have very large entanglement \cite{Karol}.
We have simulated states with a total number of 6 qubits and natural bipartition, disregarding the presence of experimental noise and the impact of finite sampling. We have analyzed 500 instances for the 1 and 2 layers case, and 200 instances for the 3, 4 and 5 layers case. The mean number of optimization steps is of the order of a few hundreds. FIG. \ref{fig:entropies} left shows the entanglement entropy computed from the trained QSVD circuit {\sl vs.} the exact entropy.

As suggested by the Solovay-Kitaev theorem, we observe fast
convergence of results for every instance we have analyzed (FIG. \ref{fig:entropies} right). The variational circuit approaches
the exact result as we increase the number of layers, whatever the entanglement is. This is related to the fact that small-depth quantum
circuits can develop large entanglement \cite{perdomo}. In this respect, it is worth mentioning that we have also analyzed Absolute Maximally Entangled (AME) states {\cite{albaame}, for which the convergence of the variational QSVD is fast and faithful. Simulations with a higher number of qubits should be carried out in the future.

\bigskip
\noindent{\sl \bf Bonus 1: SWAP without connecting gates}
A peculiar spinoff of the QSVD circuit is the possibility of performing a SWAP operation between parties $A$ and $B$ without using any gate that connects both subsystems. This is in contrast with the standard SWAP, where each pair of swapped qubits would need a series of CNOT gates.

The idea is shown in FIG. \ref{fig:swap}. It is enough to apply the QSVD to $\ket_{AB\,}$, then apply the adjoint $U^\dagger$ and $V^\dagger$ gates but acting on the opposite subsystem. That is:
\beq
(V_A^\dagger\otimes U_B^\dagger)\, (U_A\otimes V_B)\,|\psi\rangle_{AB}=|\psi\rangle_{BA} \,.
\eeq
The implementation of the adjoint unitaries only need classical communication between parties $A$ and $B$, since each unitary is characterized by a set of classical parameters.
Notice as well that none of the gates will ever cross the barrier between the two systems.

\begin{figure}[t]
	\centering
	\includegraphics[width=8.0cm,height=4.5cm]{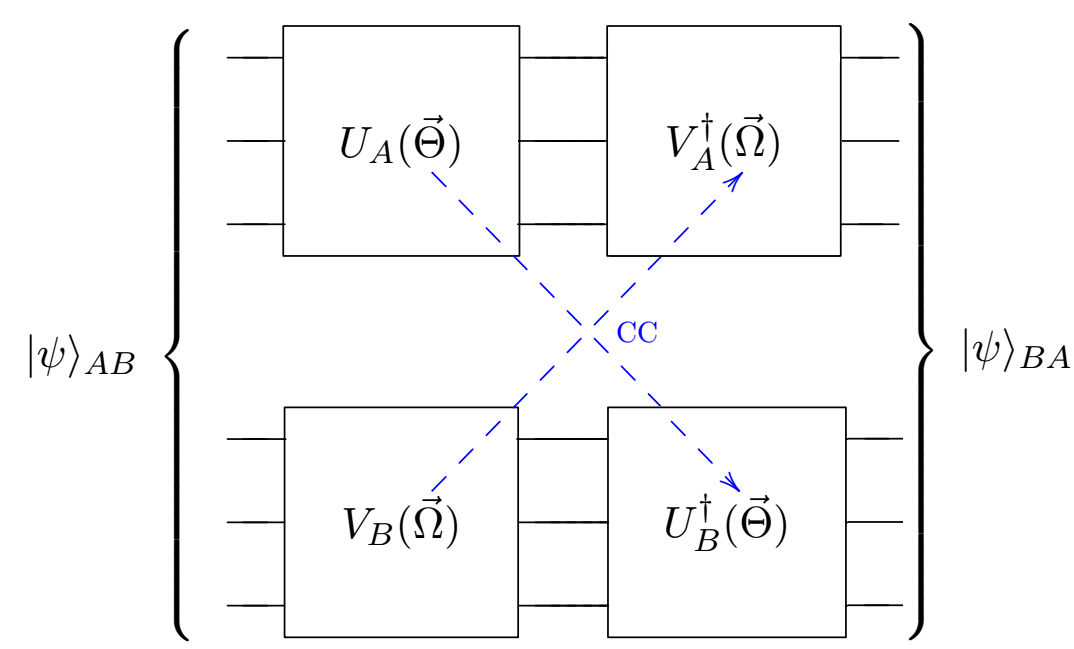}	
	\caption{Application of the QSVD followed by the adjoint $U^\dagger$ and $V^\dagger$ gates acting on opposite subsystems, mediated by classical communication ($CC$) of the optimal parameters, allows to perform a long-distance SWAP operation without the need of any quantum communication between subsystems.}
	\label{fig:swap}
\end{figure}

The possibility of performing a SWAP without quantum communication opens the possibility of swapping at long distances. Lets imagine Alice and Bob received their pieces of a given common state. They can then measure their copies and notify publicly their results. Without further communication they can improve their variational QSVD. After a number of iterations, they will observe exact coincidence. Both parties can then communicate classically the characterization of their respective unitaries, and run once more the QSVD adding the adjoint exchanged gates at the end. They will then have achieved a long distance SWAP without
quantum interaction. Of course, standard SWAP is much more powerful as it acts on a single copy of any unknown state. The price to be paid is the need for entangling gates across the two subsystems.

\bigskip
\noindent{\sl \bf Bonus 2: QSVD as a quantum encoder}
The QSVD  algorithm has a further spinoff. Let us consider for the sake of simplicity a system of $n$ qubits where we apply the QSVD algorithm to a given bipartition. 
If we consider the final state of the circuit, the exact coincidence of the parties can be used to set to $|0\ra$ all the qubits of \eg party $A$. It is only needed to apply a CNOT between each pair of coincident qubits controlled at $A$ and targeted to $B$, as illustrated in FIG. \ref{fig:encoder}. The QSVD plus a series of CNOTs corresponds to a quantum encoder designed to compress the initial state onto $(CNOT_1 \ldots CNOT_{n/2})\,(U_A\otimes V_B)\, |\psi\ra_{AB} =|00\ldots0\ra_A\, |\phi\ra_B \,,$
where
$
 |\phi\ra_B=\sum_{i=1}^\chi \lambda_i \,e^{i\alpha_i}|e_i\ra_B \,.
$
All the information of the original state has been packed into one subsystem. The circuit being unitary, this encoding can be exactly decoded back onto the total system.

\begin{figure}[t]
	\centering
	\includegraphics[width=7.8cm,height=4.45cm]{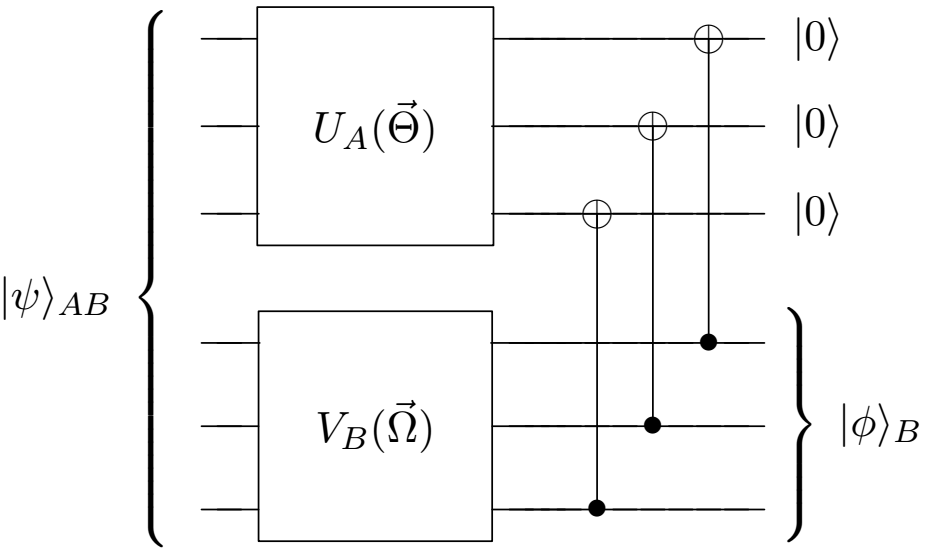}	
	\caption{ Further use of CNOT gates makes QSVD an encoder of the original quantum state $\ket_{AB}$ onto one of its parts $|\phi\ra_B$.}
	\label{fig:encoder}
\end{figure}

The same idea can be reversed. Let us imagine that we are interested in creating a random state that displays a very precise entanglement structure. The procedure would be to first manage to create the following superposition on a subsystem $|\psi\ra_A = \sum_i \lambda_i\, e^{i\alpha_i} |e_i\rangle_A$. Then a series of CNOTs connecting each qubit to an ancilla would lead to $|\psi\ra_{AB} = \sum_i \lambda_i \,e^{i\alpha_i} |e_i\rangle_A|e_i\rangle_B$. Finally, the state can be randomized by taking arbitrary unitaries on $A$ and $B$.
 
\bigskip
\noindent{\sl \bf Conclusion}
We have presented a novel algorithm, QSVD, that provides Schmidt eigenvalues and eigenvectors of any bipartite pure state, given many copies of it. Its key idea can be traced to demand {\sl exact output coincidence} on any measurement of the two parties that make the system. 

The QSVD can be used to analyze the entanglement which is present in the result of some algorithm. For instance, if a variational quantum circuit is trained to minimize the energy of {\sl e.g.} the Ising model, the final result can be run with the addition of the variational form of QSVD. The results would then allow to check the logarithmic growth of the entropy at criticality.

The QSVD seems to be a natural structure to achieve a number of quantum tasks. Here we have analyzed the possibility
to achieve a SWAP operation without any gate that connects qubits from both sides of the state. We have also shown that QSVD plus a series of CNOTs is tantamount to a quantum encoder.

\bigskip
\noindent{{\sl \bf Acknowledgements.} CBP, DGM and JIL acknowledge CaixaBank for its support of this work through Barcelona Supercomputing Center's
project CaixaBank Computación Cuántica. CBP, DGM and JIL are supported by Project PGC2018-095862-B-C22 and Quantum CAT (001-P-001644).}

\clearpage
\onecolumngrid

\appendix
\begin{center}
\Large{\bf Supplementary Material for ``Quantum Singular Value Decomposer''}
\end{center}

\twocolumngrid

\noindent {\sl \bf Variational ansatz}. The basic unit-cell or layer of the variational ansatz employed in the simulations is shown in FIG. \ref{fig:unitary}. These layers are used as buildings blocks to construct deeper circuits, by consecutively applying the architecture of the single layer, followed by a final set of single-qubit rotations. The number of layers on a circuit controls the accuracy of the estimation, as previously discussed.
  
  One layer has depth 8 (10 if the number of qubits of a subsystem is odd), so the depth of the circuit as a function of the number $l$ of layers is $8l+3$ ($10l+3$). The total number of 1-qubit gates is $6ln+3n$, where $n$ is the number of qubits, and that of 2-qubit gates is $ln$ ($ln+2l$). Therefore, depth and number of gates are efficient in both the number of qubits and the number of layers.
  
\begin{figure}[h]
  \centering
  \includegraphics[scale=0.23]{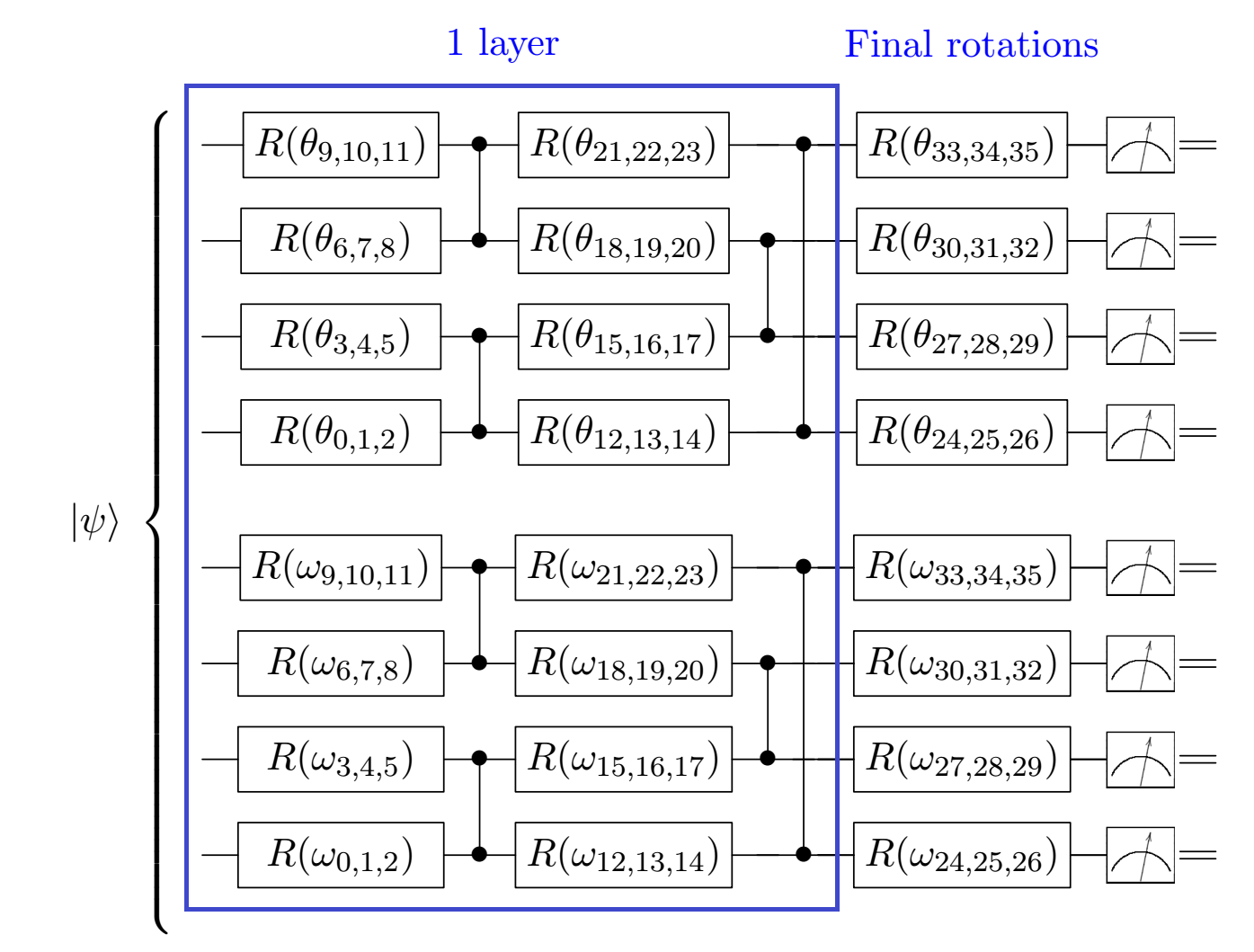}	
  \caption{Architecture of the variational circuit employed in the QSVD. Several layers of gates are applied in a consecutive manner in order to improve accuracy, and the circuit always ends with a final set of single-qubit rotations prior to measurement. The notation stands for $R(\theta_{\alpha,\beta,\gamma})\equiv Rz(\theta_\alpha)Rx(\theta_\beta)Rz(\theta_\gamma)$. If the number of qubits is odd in a given subsystem, then an extra CZ between the first and last qubit of the subsystem is added after each complete rotation.}

  \label{fig:unitary}
\end{figure}
  
\bigskip
\noindent {\sl \bf Optimization procedure}. The classical method employed in the optimization loop was L-BFGS-B, which is gradient-based and involves estimation of the inverse Hessian matrix. We utilized the implemented version of the open-source Python package SciPy Optimize \cite{scipy}.
\end{document}